\documentclass{article}

\usepackage{arxiv}

\usepackage[utf8]{inputenc} 
\usepackage[T1]{fontenc}    
\usepackage{hyperref}       
\usepackage{url}            
\usepackage{booktabs}       
\usepackage{amsfonts}       
\usepackage{nicefrac}       
\usepackage{microtype}      
\usepackage{lipsum}		
\usepackage{graphicx}
\usepackage{natbib}
\usepackage{chemformula}
\usepackage{doi}
\usepackage{amsmath} 
\usepackage{textcomp}

\title{Liquid crystal based tailoring of strong coupling  in \ch{WS2} metasurfaces: Towards reconfigurable quantum photonics}

\author{ \href{https://orcid.org/0009-0009-2803-1849}{\includegraphics[scale=0.06]{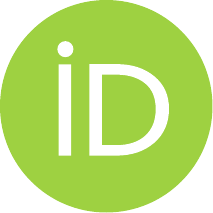}\hspace{1mm}Anu Koviloor Manian}\thanks{Use footnote for providing further
		information about author (webpage, alternative
		address)---\emph{not} for acknowledging funding agencies.} \\
	Department of Physics\\
	Mahindra University\\
	Hyderabad, 500043 \\
	\texttt{se22pphy001@mahindrauniversity.edu.in} \\
	\And
	\href{https://orcid.org/0000-0003-3428-0170}{\includegraphics[scale=0.06]{orcid.pdf}\hspace{1mm}Jayasri Dontabhaktuni} \\
	Department of Physics\\
	Mahindra University\\
	Hyderabad, 500043 \\
	\texttt{jayasri.d@mahindrauniversity.edu.in} \\
}

\renewcommand{\shorttitle}{\textit{arXiv} Template}

\hypersetup{
pdftitle={A template for the arxiv style},
pdfsubject={q-bio.NC, q-bio.QM},
pdfauthor={David S.~Hippocampus, Elias D.~Striatum},
pdfkeywords={Strong coupling, Transition metal dichalcogenides, Rabi splitting, Liquid crystals},
}

\begin{document}
\maketitle

\begin{abstract}
Strong light-matter interactions in 2D materials have garnered significant interest for their potential in nonlinear optics and quantum photonics. Transition metal dichalcogenides (TMDCs), with their robust excitonic responses, serve as promising materials for exploring these interactions. Importantly, tailoring such strong coupling giving rise to tunable quantum and non-linear emissions are under explored. In this study, we propose a novel approach that employs liquid crystals (LCs) as a tunable medium to modulate the strong coupling between TMDC excitons and photonic modes and hence gives rise to tunable quantum emissions. LCs offer high birefringence and anisotropic optical properties, which can be dynamically tuned using external stimuli such as electric fields or temperature variations. By embedding TMDCs within an LC environment and coupling them to photonic quasi-bound states in the continuum (quasi-BIC), we demonstrate precise control over the exciton-photon interactions. The orientation of the LCs, governed by applied external fields, directly influences the coupling strength, enabling real-time modulation of exciton-polariton states and Rabi-splitting. This tunability paves the way for the realization of reconfigurable and adaptive quantum photonic devices. The designed quasi-BIC metasurface facilitates enhanced light-matter interactions, with potential applications in on-chip quantum technologies, tunable polaritonic systems and active photonic components. 
Our system exhibits Rabi splitting energy of 182.5 MeV in air, and 139, 153.7, and 131 meV under different LC orientations. Additionally second order autocorrelation function calculated at zero delay yields \( g^{(2)}(0) = 0.89 \), demonstrating photon anti-bunching, confirming the quantum nature of emission. Our findings establish a versatile platform where TMDCs and LCs synergistically enable electrically controllable strong coupling, offering a scalable and responsive architecture for next-generation quantum photonic devices.
\end{abstract}


\section{Introduction}
Transition metal dichalcogenides (TMDCs) have emerged as promising quantum materials capable of efficient light absorption and emission. TMDCs exhibit strongly bound excitons with binding energies of several hundred meVs, which allows excitonic behavior to remain stable beyond room temperature \cite{chernikov2014exciton, ugeda2014giant, he2014tightly}. High refractive index, weak van der Waals forces, and atomically confined excitons even in the bulk makes them a compelling alternative for conventional semiconductors \cite{arora2017interlayer, weber2023intrinsic}. These materials also offer high mechanical stability, large oscillator strength, and valley degree of freedom, making them ideal for room temperature polaritonic and quantum photonic devices \cite{sun2017optical,chen2017valley}. A significant achievement is the emergence of hybrid light-matter quasi-particles known as exciton-polaritons, resulting from the strong coupling between excitons and confined optical modes. Coherent and reversible energy exchange between excitons and photons occurs when coupling strength exceeds system's total dissipation. Such hybrid states facilitate phenomena such as Rabi splitting, low-power optical switching, Bose-Einstein condensation, non-linear quantum optics, and polariton lasing \cite{su2017room,amo2009superfluidity, deng2010exciton,amo2010exciton}. While early studies focused on monolayers of TMDCs \cite{chernikov2014excitons,gan2013controlling}, due to their pronounced photoluminescence and direct bandgap, recent works demonstrate that multilayer\cite{qin2023strong} and bulk TMDCs\cite{munkhbat2018self} also serve as intriguing platforms for robust light-matter coupling.

Bulk TMDCs have an indirect bandgap, which renders light emission less effective than in a monolayer\cite{wang2012electronics}. However, the strong exciton absorption peak and high refractive index of bulk TMDCs enable the spatial overlap of excitons and the optical resonance mode inside the same structure, making them a viable option for enhancing strong coupling \cite{wang2022boosting, verre2019transition}. Strong light-matter coupling incorporating TMDC materials was attained in a variety of cavity-emitter combinations such as single plasmonic nanoparticles \cite{wen2017room,stuhrenberg2018strong,cuadra2018observation,zheng2017manipulating,kleemann2017strong}, high quality factor dielectric cavities\cite{dufferwiel2015exciton,sun2017optical,chen2017valley}, or diffractive nanoparticle arrays \cite{liu2016strong}. All of the aforementioned research have strong coupling obtained by the use of well-defined resonant optical modes supported by independent resonators. Mode splitting in the absorption spectra\cite{munkhbat2018self, savona1995quantum, zengin2016evaluating,antosiewicz2014plasmon} is interpreted as Rabi splitting in the strong coupling domain. This facilitates quantum coherent oscillations between the coupled systems and the quantum superpositions of distinct quantum states, which are fundamentally crucial for quantum information processing \cite{hennessy2007quantum,peter2005exciton,reithmaier2004strong}. Consequently, Rabi splitting has been thoroughly examined in several quantum and semi-classical systems, including quantum-dot-microcavity \cite{hennessy2007quantum,peter2005exciton,reithmaier2004strong} and emitter-plasmon systems \cite{chikkaraddy2016single,schlather2013near,gomez2010surface}, with investigations encompassing emission as well as scattering, transmission, reflection and absorption characteristics.
A variety of tuning mechanisms are being employed to tailor the strong coupling which includes incident angle\cite{munkhbat2018self,do2024room}, geometry\cite{zong2022enhanced, zhang2020hybrid}, temperature \cite{wen2017room}, lattice constant \cite{zong2022enhanced,qin2023strong,zhang2020hybrid}, 
 and polarization configuration\cite{zong2022enhanced}. Although liquid crystals and TMDCs have made tremendous strides in recent years, the possibility of combining the two to create a tunable strong coupling mechanism in which the interaction between matter and light can be dynamically controlled remains unexplored.

 \begin{figure}[ht]
\centering
\includegraphics[width=\linewidth]{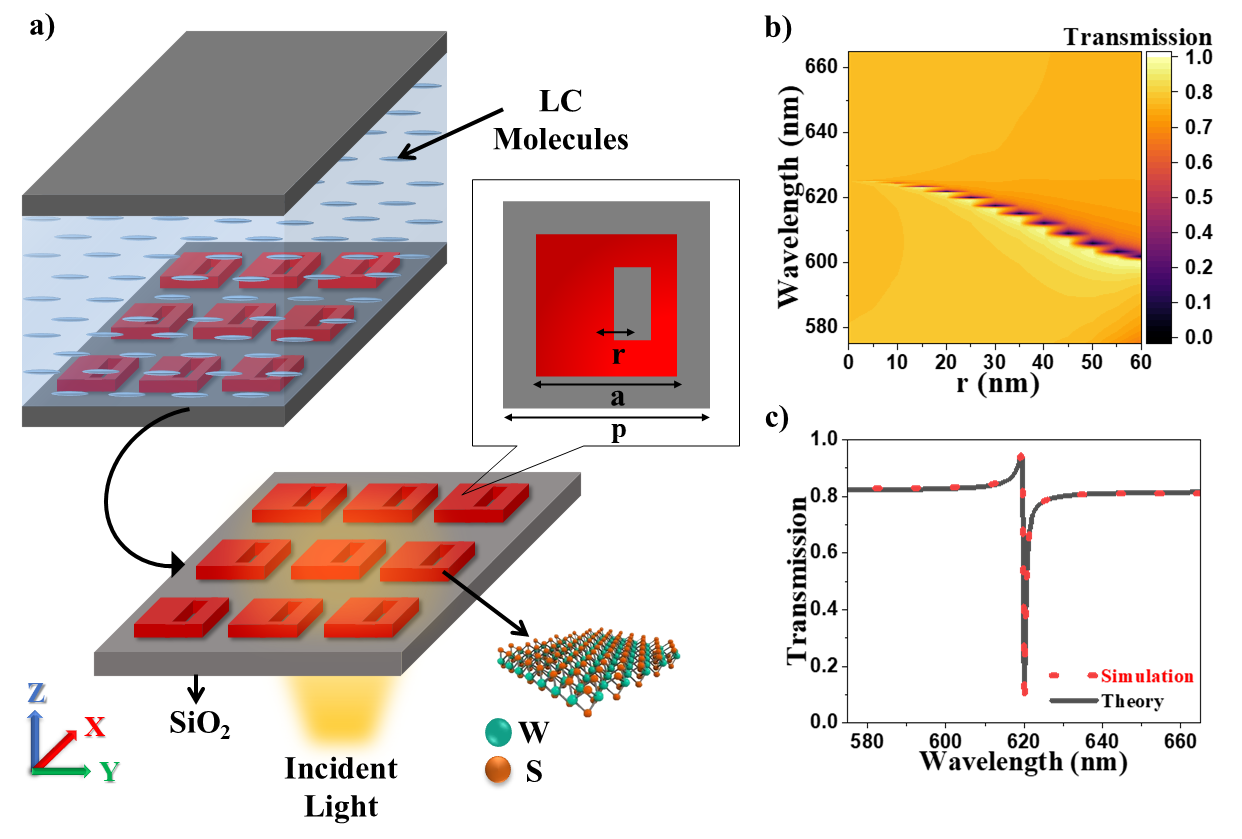}
\caption{(a) Schematic representation bulk  \ch{WS2} metasurface in liquid crystal medium with asymmetrical rectangular holes arranged in SiO$_2$ substrate which is illuminated by a plane wave. (b) Transmission contour plot with respect to variation in the asymmetry parameter r. (c) The transmission curve of uncoupled quasi-BIC corresponding to theoretical and numerical simulation, respectively.}
\label{fig1}
\end{figure}

 Liquid crystal (LC) is a configurable platform with high optical anisotropy and birefringence which supports programmable, multifunctional, and tunable electromagnetic response\cite{shabanpour2021real} in the metasurfaces through thermal, optical, electrical and magnetic tuning \cite{rocco2022tunable, sautter2015liquid, saifullah2022recent, komar2017electrically, izdebskaya2022magnetic}. The applied electric field causes LC molecules to reposition their optic axis, which in turn causes the incident light's polarization direction to change. The direction of polarization of incident light can therefore be adjusted in relation to the external electric field to alter the electromagnetic response of structures enclosed in the LC. Some applications of metasurfaces embedded in liquid crystals include metalens\cite{lininger2020optical}, ultrafast switching\cite{sakhare2022ultra}, varifocal lenses\cite{bosch2021electrically}, tunable second harmonic generation\cite{rocco2022tunable}, spatial light modulators\cite{li2019phase}, color filters\cite{xie2017liquid}, etc. Works are already done to show impact of electrically driven anisotropic LC media on adjusting the anapole states, creating new opportunities for programmable non-radiative mode applications \cite{sakhare2023tailoring}. Majority of the literature on LC-based metasurfaces concentrates on the external field-dependent switching between the resonant modes and the tunability of resonance frequencies as a function of LC orientations. Notwithstanding advancements in recent decades, achieving active control of strong coupling remains challenging due to the restricted tunability of excitons in molecules or quantum dots. Furthermore, customizing the properties of a metasurface with a fixed shape and composition is difficult. However, in order to push the interactions down to the single exciton level, carefully crafted architectures must be used, which significantly reduces the robustness of the strong coupling. Thus far, no one has worked on using LCs to control or adjust the strong coupling mechanism.

 Here, for the first time, we are leveraging LCs to tune the strong coupling interaction associated with a bulk TMDC system. Our design consists of a \ch{WS2} metasurface having a asymmetrically placed rectangular hole inside and arranged in SiO$_2$ substrate which produces high Rabi splitting. The strong coupling can be characterized by the considerable Rabi splitting of 182.5 meV when the system is in air medium and 139, 153.7 and 131 meV in the LC medium. The absorption spectrum's anti-crossing characteristic serve as additional evidence of this strong coupling between exciton and polariton. The system has been studied in the air medium as well as in LC and explored the modes associated for the enhancement of strong coupling. The absorption spectrum has been examined with respect to effect of in periodicity in the case of each orientation of LC and also with incident angle. We have also shown the correlation function at zero delay having a value of 0.89, indicates photon antibunching which is crucial for system to exhibit quantum behavior. This opens up a new field of study since LCs dynamic, responsive nature may make it possible to control or modify the behavior and strength of coupled systems. Our findings hold considerable promise for controlling exciton-polariton at room temperature and offer a viable option for photonic devices that use strong coupling phenomenon.

\begin{figure}[ht]
\centering
\includegraphics[width=\linewidth]{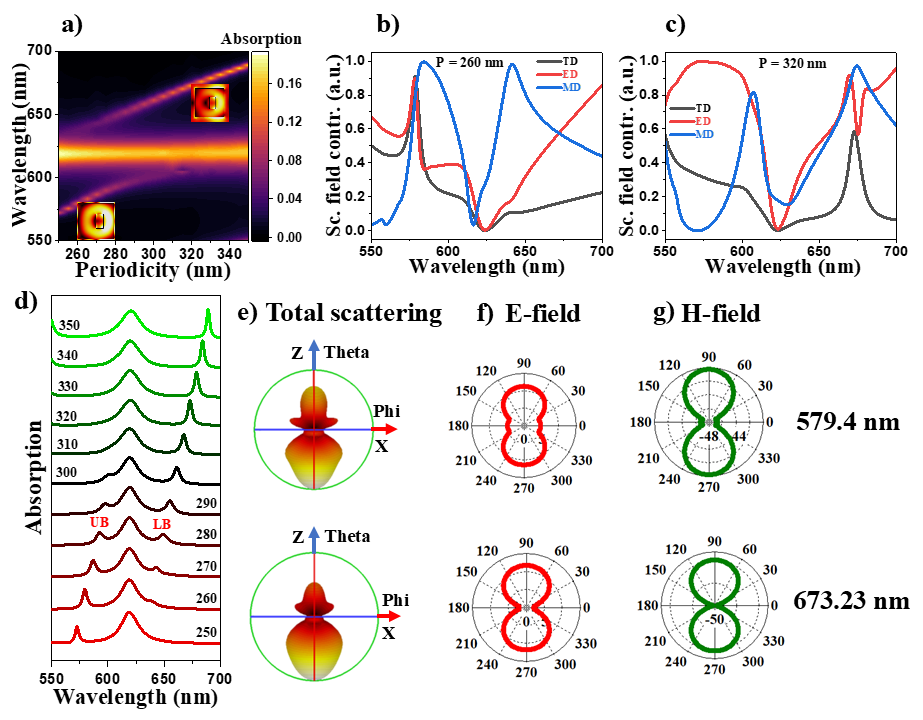}
\caption{Periodicity dependence in the air medium. (a) Contour plot of absorption spectra with respect to periodicity varying from 250 to 350 nm in the wavelength range of 550 to 700 nm along with corresponding near-field profiles of lower and upper branches. (b, c) The scattered field contributions corresponding to the periodicity 260 nm and 320 nm. (d) Absorption spectra of bulk \ch{WS2} metasurface for different periodicity's indicating the Rabi splitting with lower and upper branch. (e - g) Far-field radiation profiles showing total scattering, E-field and H-field at 579.4 nm and 673.23 nm.}
\label{fig2}
\end{figure}

\begin{figure}[ht]
\centering
\includegraphics[width=\linewidth]{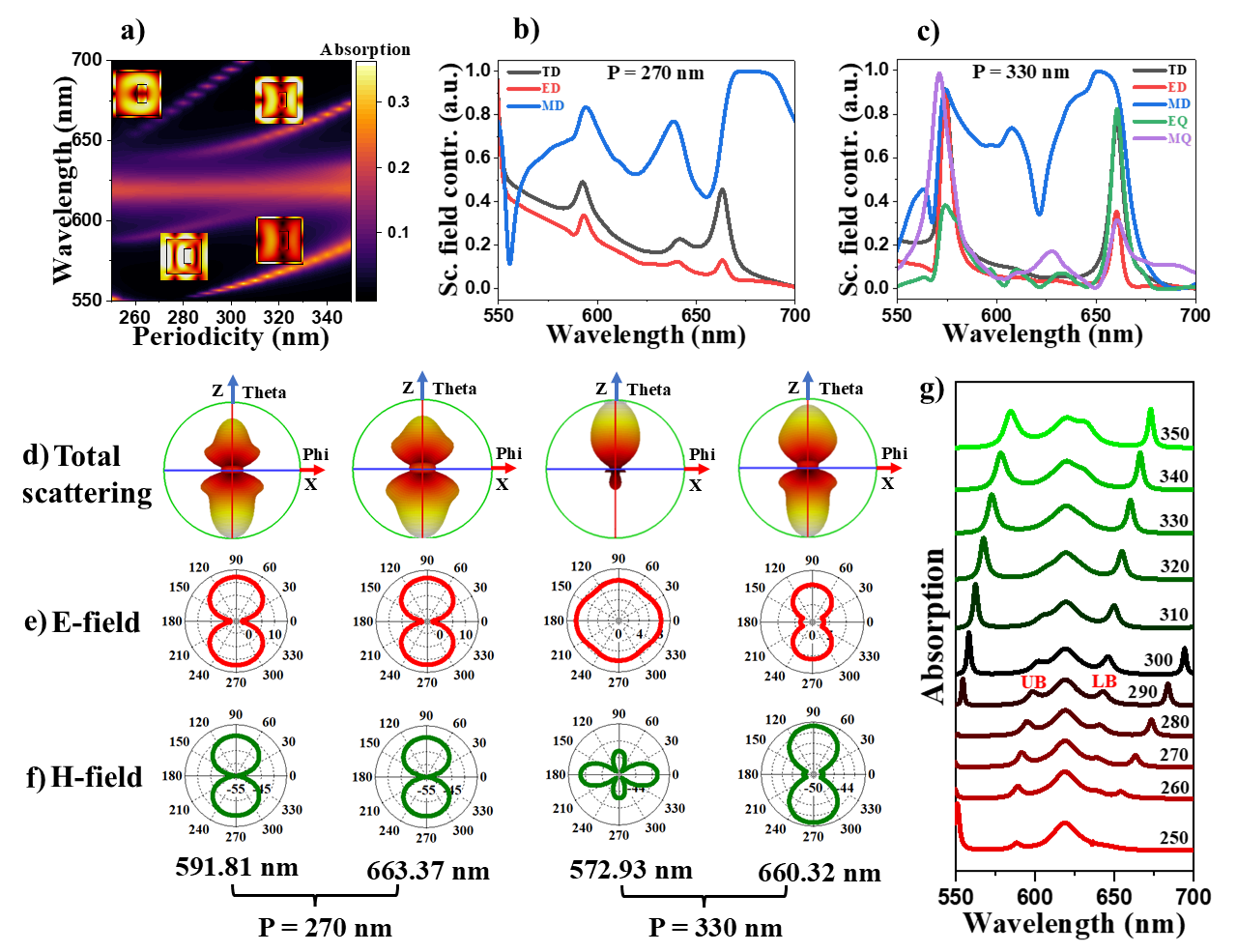}
\caption{LC along X-axis. (a) Contour plot of absorption spectra with respect to periodicity varying from 250 to 350 nm in the wavelength range of 550 to 700 nm along with corresponding near-field profiles of lower and upper branches. (b, c) The scattered field contributions corresponding to the periodicity 270 nm and 330 nm. (d - f) Far-field radiation profiles showing total scattering, E-field and H-field at 591.81 nm, 663.37 nm, 572.93 nm and 660.32 nm. (g) The absorption spectra indicating the Rabi splitting.}
\label{fig3}
\end{figure}

\section{Results and Discussion}
\subsection{Effect of periodicity in air medium}
A TMDC metasurface with asymmetrically positioned rectangular holes on \ch{WS2} nanocuboids in silica substrate is displayed in the Figure \ref{fig1}(a). The asymmetry parameter is denoted by r = 25 nm is the distance from center of nanocuboid to center of rectangular hole. Figure \ref{fig1}(b) shows the contour plot of transmission spectra with respect to asymmetry parameter (r) within the periodicity (P) 280 nm. For the designed \ch{WS2} metasurface, an ideal BIC with vanished linewidth at around 625 nm is observed. The Fano formula fits the sharp resonance well, demonstrating that the ideal BIC is converted to quasi-BIC by breaking the unit cell's in-plane symmetry through energy exchange and shown in \ref{fig1}c). The P dependence in the metasurface has been studied by numerical simulations based on the finite element method (FEM) using commercially available CST software, as explained in the (\textit{Methods}) section. Figure \ref{fig2} indicates how periodicity variation is effective in the strong coupling mechanism with respect to the modes exhibited by the \ch{WS2}, when the system is in the air medium. Figure \ref{fig2}(a) shows anti-crossing behavior in the absorption spectra along 550 to 700 nm when the P is varied from 250 to 350 nm accompanied with respective near-field profiles. To study the modes associated with the lower and upper branches, multipole analysis has been done and Figure \ref{fig2}(b, c) shows the corresponding results for the case in which P = 260 nm and P = 320 nm. The lower branch depicts the domination of magnetic dipole (MD) where the toroidal dipole (TD) and electric dipole (ED) undergoes constructive interference. At P = 320 nm, where the upper branch shows the MD with the TD and ED forming destructive interference. The far-field radiation results showing total scattering, E-field and H-field is given in Figure \ref{fig2}(e-g). 579.4 is corresponding to P 260 nm and 673.23 nm is corresponding to P = 320 nm representing the lower and upper branches shows bidirectional scattering along with ED and MD.

The Rabi splitting occurs in the strong coupling domain, when the interaction causes the creation of polariton modes. The strong coupling between the exciton and quasi-BIC modes are described by the coupling Hamiltonian as \cite{huang2022enhanced, qin2023strong, zong2022enhanced}, 
\begin{equation}
H=\begin{pmatrix}
E_{q-BIC} + i\gamma_{q-BIC} &  g \\
g & E_{ex} + i\gamma_{ex} 
\end{pmatrix}
\end{equation}
 where g is the coupling strength between BIC and exciton mode and E$_{q-BIC}$ and E$_{ex}$ are the energy of the uncoupled quasi-BIC and exciton modes and, $\gamma_{q-BIC}$ and $\gamma_{ex}$ are the decay rate of the uncoupled quasi-BIC and exciton modes. The diagonalization of the Hamiltonian yields the solution as follows, 
\begin{equation}
E^{\pm} = \frac{E_{q-BIC} + E_{ex}}{2} + i\frac{\gamma_{q-BIC} + \gamma_{ex}}{2} \pm \sqrt{g^2 + \frac{1}{4}[(E_{q-BIC} - E_{ex})+i(\gamma_{ex} - \gamma_{q-BIC})]^2}
\end{equation}
At zero detuning, E$_{q-BIC}$ - E$_{ex}$ = 0 and thus,
\begin{equation}
E^{\pm} = \frac{E_{q-BIC} + E_{ex}}{2} + i\frac{\gamma_{q-BIC} + \gamma_{ex}}{2} \pm \sqrt{g^2 - \frac{1}{4}[(\gamma_{ex} - \gamma_{q-BIC})]^2}
\end{equation}
This energy difference is denoted as the Rabi splitting and given by, 
\begin{equation}
E^+ - E^- = \sqrt{4g^2 - (\gamma_{ex} - \gamma_{q-BIC})^2}
\end{equation}
Thus Rabi splitting ($\hbar\Omega_{R}$) can be calculated as $E^+ - E^- = \hbar\Omega_{UB}-\hbar\Omega_{LB}$ where, $\hbar\Omega_{LB}$ and $\hbar\Omega_{UB}$ are the energies corresponding to the lower and upper branches respectively. At periodicity of 280 nm, the exciton and quasi-BIC mode are nearly resonant, resulting in zero detuning. Considering the absorption corresponding to the periodicity 280 nm in the Figure \ref{fig2}(d) having the upper branch (UB) is at around 592.4 nm and the lower branch (LB) is at around 649 nm. The calculated Rabi splitting is 182.5 meV  which is larger than the conventional Rabi splitting formed in bulk TMDC \cite{qin2023strong,zong2022enhanced}. In addition to the anti-crossing in the absorption spectrum, strong coupling can be verified by satisfying the following condition,
\begin{equation}
\hbar\Omega_{R} > \gamma_{ex} + \gamma_{q-BIC} 
\end{equation}
Our system has $\gamma_{ex}$ = 43.6 meV and $\gamma_{q-BIC}$ = 2.89 meV which rigorously meets this criterion with a Rabi splitting of 182.5 meV. Further, to demonstrated the quantum characteristic of the designed metasurface using the correlation function at zero delay (\( g^{(2)}(0) \)) as shown in (\textit{Methods}) section. The framework computes \( g^{(2)}(0) = 0.89 \), a value that implies photon antibunching which open ups the persistent influence of quantum interactions.

\begin{figure}[ht]
\centering
\includegraphics[width=\linewidth]{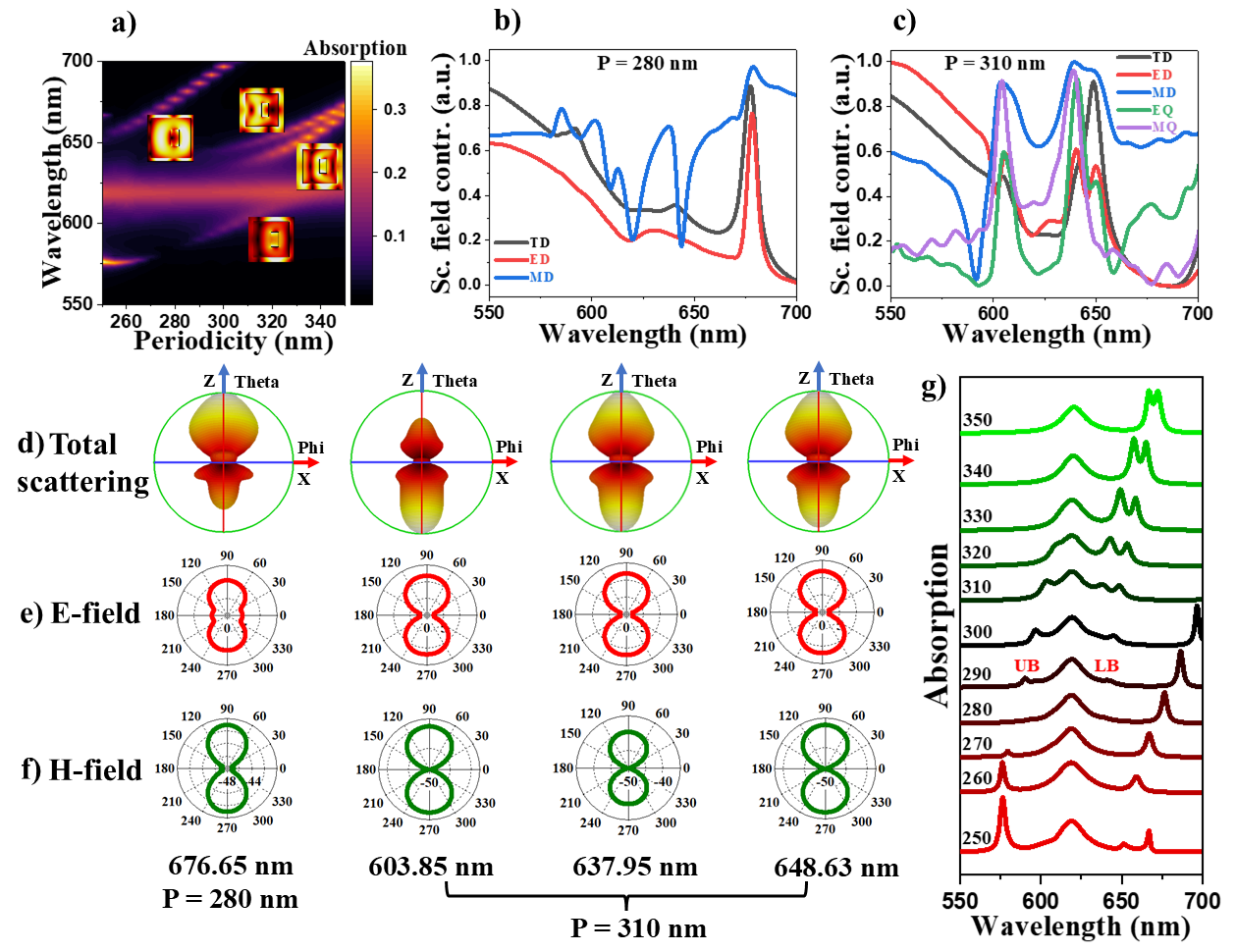}
\caption{LC along Y-axis. (a) Contour plot of absorption spectra with respect to periodicity varying from 250 to 350 nm in the wavelength range of 550 to 700 nm along with corresponding near-field profiles of lower and upper branches. (b, c) The scattered field contributions corresponding to the periodicity 280 nm and 310 nm. (d - f) Far-field radiation profiles showing total scattering, E-field and H-field at 676.65 nm, 603.85 nm, 637.95 nm and 648.63 nm. (g) Absorption spectra marked with LB and UB indicating the Rabi splitting.}
\label{fig4}
\end{figure}

\subsection{Effect of periodicity in LC orientation}
The \ch{WS2} metasurface has been studied under LC medium. Figure \ref{fig3} refers to the results obtained when the LC placed along the X-axis. In the absorption spectra with respect to P, Figure \ref{fig3}(a) indicates anti-crossing behavior along with an upper and lower splitting in the spectra. Figure \ref{fig3}(b) showing the multipole study at the periodicity of 270 nm depicting the modes corresponding to the field profiles for UB (591.81 nm) and additional splitting (663.37 nm) happening in P range 250 to 300 nm along the wavelength 650 to 700 nm. The UB occurs due to the MD contribution, where TD and ED undergoes constructive interference. MD have high contribution at 663.37 nm also along with the constructive interference formed by contributions of TD and ED. Another additional splitting is happening from 290 to 650 nm along 550 - 600 nm wavelength range. Figure \ref{fig3}(c) at P 330 nm corresponds to a higher contribution of magnetic quadrupole (MQ) moment along with domination of TD, ED and MD at 572.93 nm which corresponds to a peak for the extra splitting happened in the lower wavelength regime. The LB (660.32 nm) is MD with equivalent contribution of TD and MQ. The far-field results shows bi-directional scattering along the anti-crossing branches representing UB and LB along with ED and MD. The additional splitting happening in higher wavelength region also follows the same far-field results. Whereas the splitting in the lower wavelength region represents forward scattering along with electric and magnetic field forming EQ and MQ. The calculated Rabi splitting is 139 meV by taking the absorption corresponding to the periodicity 290 nm at which the exciton and quasi-BIC mode are nearly resonant with UB and LB around 602.46 nm and 646.24 nm (Figure \ref{fig2}(g)).

\begin{figure}[ht]
\centering
\includegraphics[width=\linewidth]{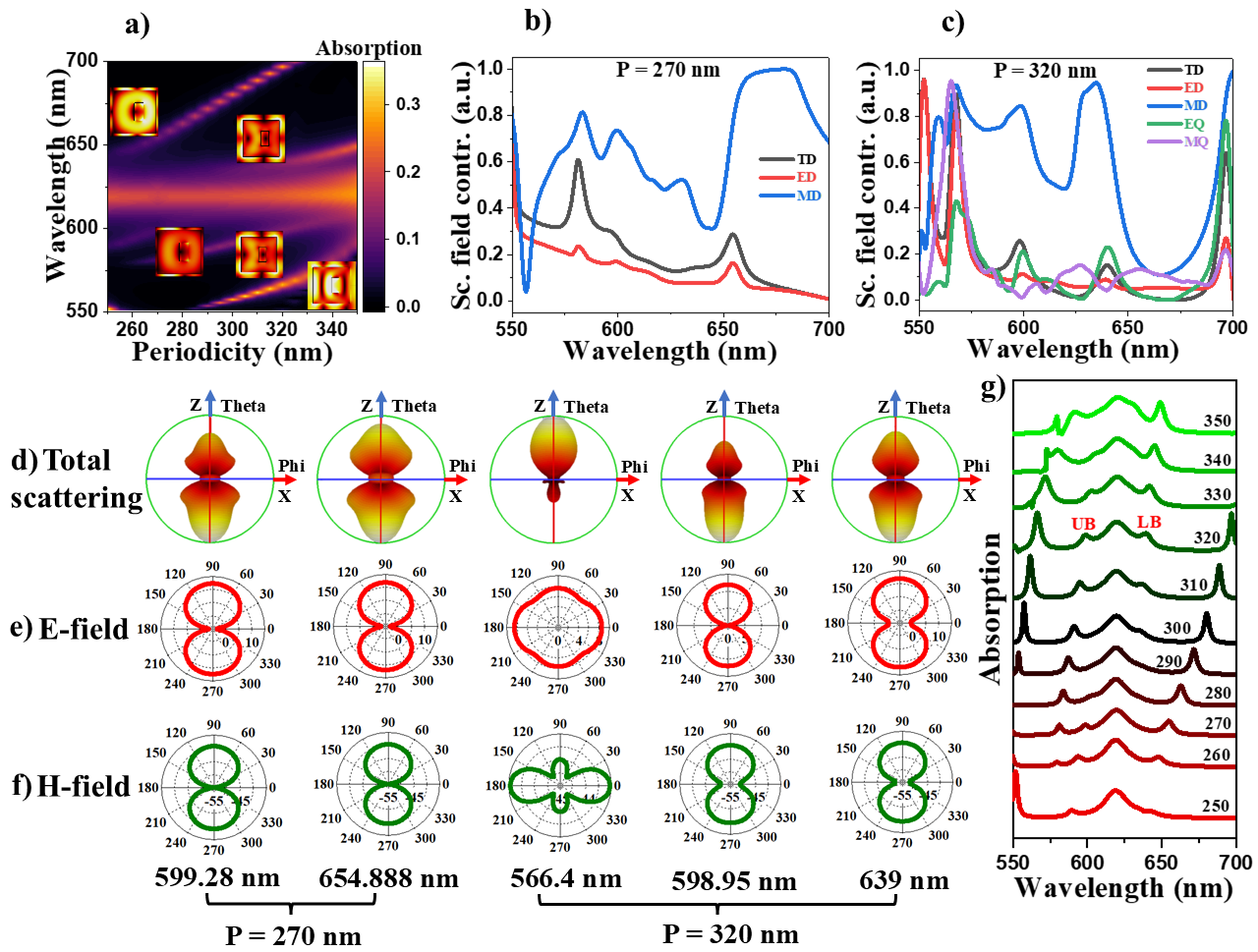}
\caption{LC along Z-axis. (a) Contour plot of absorption spectra with respect to periodicity varying from 250 to 350 nm in the wavelength range of 550 to 700 nm along with corresponding near-field profiles of lower and upper branches. (b, c) The scattered field contributions corresponding to the periodicity 270 nm and 330 nm. (d - f) Far-field radiation profiles showing total scattering, E-field and H-field at 591.81 nm, 663.37 nm, 572.93 nm and 660.32 nm. (g) Absorption spectra marked with LB and UB indicating the Rabi splitting.}
\label{fig5}
\end{figure}

\begin{figure}[ht]
\centering
\includegraphics[width=\linewidth]{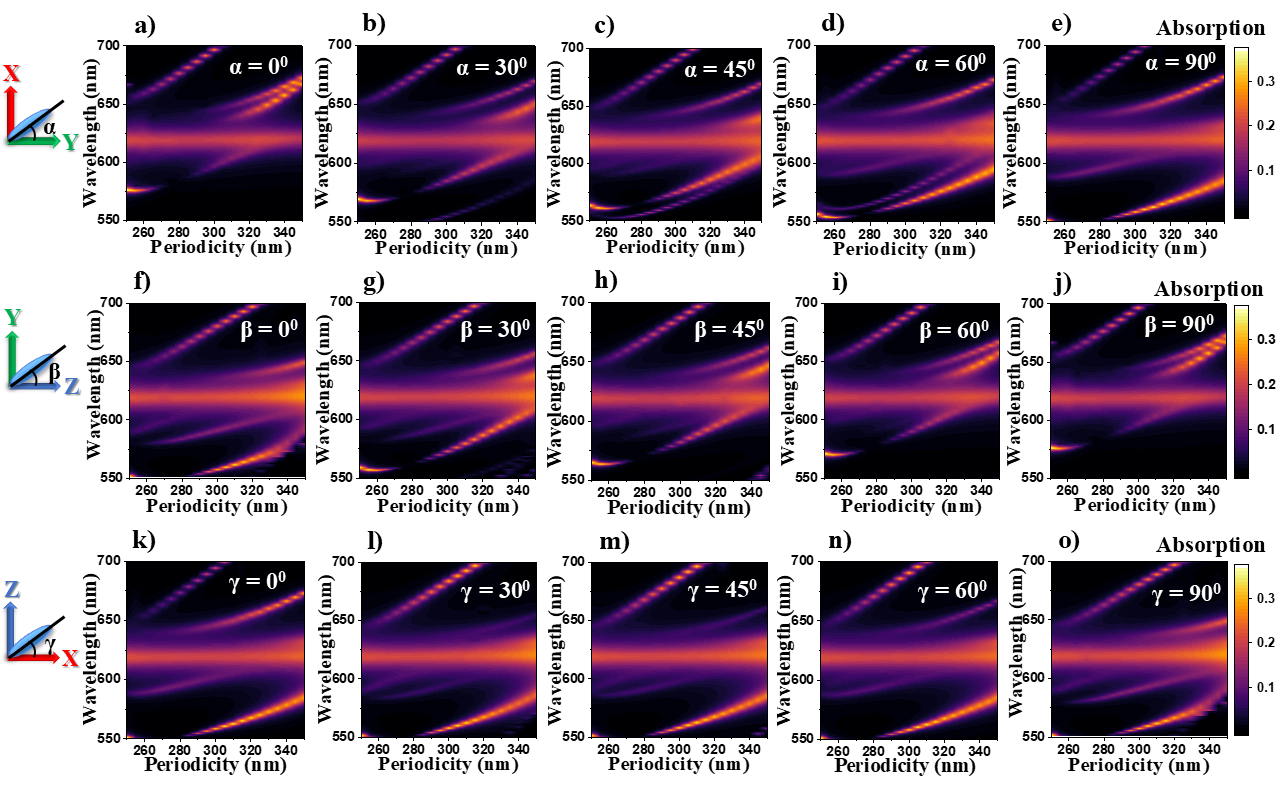}
\caption{Effect of LC orientations in the absorption spectra. (a - e) Absorption spectra observed when LC orientation is along XY-plane with $\alpha$ varying from 0$^0$ to 90$^0$. (f - j) Absorption spectra observed when LC orientation is along YZ-plane with $\beta$ varying from 0$^0$ to 90$^0$ and (k - o) represents when LC orientation is along ZX-plane with $\gamma$ varying from 0$^0$ to 90$^0$.}
\label{fig6}
\end{figure}

Figure \ref{fig4} corresponds to the condition in which \ch{WS2} metasurface along with LC has been placed along Y-axis. The contour plot of absorption spectra (Figure \ref{fig4}a) represents the anti-crossing of the exciton and the polariton along with a UB and LB in the P variation of 280 - 350 nm. The modes has been analyzed along this regime and Figure \ref{fig4}c) corresponds to the mode analysis at P 310 nm. The UB is observed with contributions from MD and MQ whereas, LB which undergoes splitting and MD contributes more. MD contributed LB which is closer to the exciton has higher contributions from EQ and MQ also. The other splitting has contributions from TD also. There is branching happening in the 650 to 700 nm regime before the strong coupling happens and this is analyzed with MD (Figure \ref{fig4}b) with TD and ED forming constructive interference. The corresponding far-field results (Figure \ref{fig4}d-f) along the branches refers to bi-directional scattering and with E-field and H-field resulting in ED and MD. Figure \ref{fig4}g) represents the absorption spectra plotted with respect to P. At P = 290 nm, where the exciton and quasi-BIC mode are nearly resonant, the Rabi splitting has been calculated. UB is at around 596.93 nm and LB is around 644.63 nm and the Rabi splitting energy obtained is around 153.7 meV.

\begin{figure}[htpb]
\centering
\includegraphics[width=\linewidth]{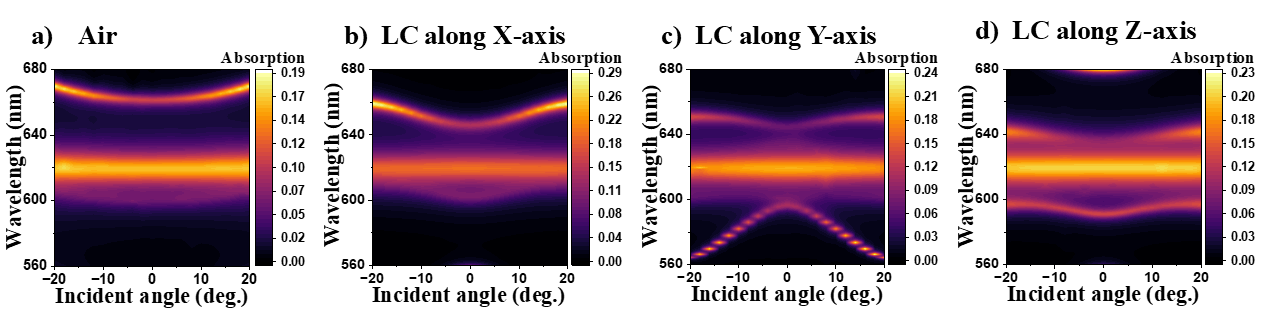}
\caption{Angle-resolved absorption spectra. (a) Contour plot absorption with respect to incident angle when the \ch{WS2} metasurface is in air medium along 560 - 680 nm wavelength region. Contour plots absorption with respect to incident angle in the case of LC along X-axis (b), Y-axis (c) and Z-axis (d) respectively.}
\label{fig7}
\end{figure}

Figure \ref{fig5} represents the case in which the \ch{WS2} metasurface has been studied under LC along along Z-axis. The contour plot of absorption spectra (Figure \ref{fig5}a) including the near-field profiles in each branch represent a clear anti-crossing behavior along the P variation from 280 to 350 nm. There are three further branchings happening in the in the case of LC along Z-axis across the same wavelength regime where the system has already been studied. Figure \ref{fig5}b) shows the mode analysis at P 270 nm representing the branches before the strong coupling. The low absorption branching happening between the P 250 to 280 nm and the branching between 650 - 700 nm is due to MD contribution. In the strong coupling region, an example of mode analysis at P 320 nm (Figure \ref{fig5}c) has been taken and UB and LB are due to contribution from MD. There is an extra branching happening below the UB which has contributions from TD, ED, MD and MQ. The far-field results (Figure \ref{fig5}d-f) shows bi-directional scattering with E-field and H-field forming ED and MD in the case of UB, LB and for the branchings happening before strong coupling regime. The branching happening below the UB has uni-directional scattering along with EQ and MQ. In order to calculate the Rabi splitting, an absorption spectrum has been plotted with respect to P as shown in Figure \ref{fig5}g). Considering the absorption corresponding to P 320 nm, the UB is observed at around 598.66 nm and LB is at around 639.25 nm. Thus the calculated value of Rabi splitting in this case is 131 meV.

Figure \ref{fig6} refers to the case in which the strong coupling has been tailored by varying the orientation of LC from $0^0$ to $90^0$ along XY, YZ and ZX plane. The polarization direction of the incoming electromagnetic wave impinge on the metasurface varies with the change in orientations of the LC in response to the external field. The alteration in polarization direction via the metasurface influences exciton-photon coupling. We have already shown the exciton-photon interactions leading to the lower and upper polaritonic levels in the air medium. As LC introduced into our system, additional level splittings are also observed along with the LB and UB. We further studied the effect of in-plane and out-plane orientations of LC in strong coupling. Figure \ref{fig6}a-e) shows the absorption spectra when LC is placed along XY-plane (in-plane) and angle of orientation ($\alpha$) varied by $0^0$, $30^0$, $45^0$, $60^0$ and $90^0$. In this case, $\alpha = 0^0$ and $\alpha = 90^0$ refers to LC along Y-axis and X-axis. The anti-crossing behavior in the absorption spectra is visible in each orientation of $\alpha$ even though LC is inducing further splittings with in the wavelength range and P variation under study. Figure \ref{fig6}f-j) corresponds to the absorption spectra in the case of YZ-plane (out-plane). The angle of orientation in this case is $\beta$ which is varied from $0^0$ (LC along Z-axis) to $90^0$ (LC along Y-axis). Each orientation of $\beta$ can can also tailor the exciton-photon coupling which is clearly visible in the Figure \ref{fig6}f-j). Similar kind of outcomes has been observed in the case of ZX-plane (out-plane) orientations also and shown in Figure \ref{fig6}h-o). Where, $\gamma$ is the angle of orientation of LC molecules along ZX-plane with $\gamma = 0^0$ and $\gamma = 90^0$ representing the case of LC along X-axis and Z-axis respectively. This inferences clearly suggest that the in-plane and out-off-plane orientations of LC can also tune the exciton-photon coupling. This method put forward different ways of tuning the the exciton-photon coupling studied with the help of periodicity dependence in absorption spectra just by varying the orientation of LC molecules. This study thus opens up a new technique of tuning the strong coupling mechanism occurring in TMDC metasurfaces.

We have also studied the effect of incident angle in the designed \ch{WS2} metasurface and shown in the Figure \ref{fig7}. In this study, we considered the \ch{WS2} metasurface at P = 300 nm and angle of incidence $\theta$ is varied from -20\,\textdegree to  20\,\textdegree. The absorption plot in Figure \ref{fig7} a) refers to the case in which the metasurface is under air medium. The exciton resonance (around 620 nm) shows a dominant absorption which is angle independent. Notwithstanding the structural asymmetry of the metasurface, there is an angular symmetry in the absorption. Figure \ref{fig7} (b-d) corresponds to the case when the metasurface is placed in LC medium along X-, Y- and Z-axis. The exciton resonance remains angle independent where as there are mode splittings which are varying with respect to the LC orientations. Absorption spectra when LC along Y-axis (Figure \ref{fig7} (c)) exhibits significant angle-dependent splitting below the exciton wavelength which can further enhance angle-sensitive coupling.

\section{Conclusion}
Through this study, we theoretically investigated the strong coupling between \ch{WS2} exciton and quasi-BIC modes. The Rabi splitting energy up to 182.5 meV is observed in the absorption spectrum of the bulk \ch{WS2} structure in air medium. In LC medium, Rabi splitting energies of 139, 158.7, and 131 meV is obtained for LC along X, Y, and Z axis. Additionally, anti-crossing behavior which is a defining element of strong coupling can be achieved by tuning the periodicity. Furthermore, the novelty of this study illustrates the efficacy of employing LCs to manipulate the strong coupling mechanism in bulk \ch{WS2} metasurface. Through the integration of LC medium, we attained adjustable control over the exciton-quasi-BIC coupling strength, providing a versatile method for dynamically modulating light-matter interactions within \ch{WS2}. Our findings demonstrate that LC-based tuning offers a precise and reversible approach to modulate excitonic characteristics without necessitating invasive mechanical or thermal alterations, rendering it a compelling tool for sophisticated optoelectronic applications. Our results provides a robust foundation for developing tunable exciton-polariton devices. The correlation function at zero delay is calculated to be \( g^{(2)}(0) = 0.89 \), a value that signals photon antibunching. This subtle deviation from classical coherence underscores the persistent influence of quantum interactions which is crucial to engineer metasurfaces for quantum applications.


\section{Methods}
\subsection{Numerical simulation}
We modeled the system using finite element method (FEM) simulations in commercially available CST software. Under the illumination of a typically incident y-polarized plane wave, a nanocuboid bulk \ch{WS2} metasurface is integrated on a SiO$_2$ substrate with a thickness of $t = 100$ nm. The lattice constant has been varied from 250 to 350 nm and each nanocuboids has sides $a = 240$ nm with a height $h = 30$ nm, respectively. A rectangular hole ($100 \times 50$ nm) at a fixed distance $r = 25$ nm from the center of \ch{WS2} to the center of the rectangular hole to make the system asymmetric. The system has been examined both in the air medium and liquid crystal medium. Unit cell simulation using periodic boundary condition (PBC) along $X$ and $Y$ directions yields the results. Simulations are implemented by referencing the dielectric function of bulk \ch{WS2} \cite{munkhbat2018self,qin2023strong,zong2022enhanced} given by,

\begin{equation}
\varepsilon(\omega) = \varepsilon_0 + \frac{f_0 \omega^2_{ex}}{\omega^2_{ex}-\omega^2-i\gamma_{ex}\omega}
\end{equation}
where $\varepsilon_0 = 20$ is the background permittivity, $f_0 = 0.2$ is the oscillator strength, $\omega_{ex}=2$ eV and $\gamma_e{}_x = 50$ meV is the exciton full-width. $f_0$ is set to be zero for background index-only material. 

High birefringence LC (1825)\cite{kowerdziej2012tunable,sakhare2023tailoring,da2009voltage} in the nematic phase with dielectric permittivity $\Delta\varepsilon = 1.44$ is used in the simulation technique. The director axes can be reoriented by the external field when all molecules aligned in the Y-direction and the LC optic axis lie in the YZ plane. This will be in the form of $n =\begin{Bmatrix}
0, \cos\beta, \sin\beta
\end{Bmatrix}$, where $\beta$ is the angle between +Y-axis and the director. The permittivity tensor used for the description of LC molecules is given by,
\begin{equation}
\varepsilon(LC)=\begin{pmatrix}
\varepsilon_{\bot } & 0 & 0\\
0 & \varepsilon_{\bot }+\varepsilon_a\cos^2\beta & \varepsilon_a\sin\beta\cos\beta\\
0 & \varepsilon_a\sin\beta\cos\beta & \varepsilon_{\bot }+\varepsilon_a\sin^2\beta 
\end{pmatrix}
\end{equation}
In the matrix mentioned above, $\varepsilon_a = \varepsilon_{\parallel} - \varepsilon_{\bot }$. Where, $\varepsilon_{\parallel} = n^2_e$ and $\varepsilon_{\bot } = n^2_o$ are the parallel and perpendicular permittivity with $n_o$ and $n_e$ being the ordinary and extraordinary refractive indices of the LC.

\subsection{Multipole analysis}
To analyze the electromagnetic response derived from the scattering parameters, multipole analysis of the scattered radiation is carried out. To differentiate the contributions originating from toroidal (\textbf{T}) and electric dipole moments (\textbf{p}), the analysis was conducted using the Cartesian coordinate system.
The equation \cite{evlyukhin2016optical} provides the multipole contributions to the total dispersed radiation from the total electric dipole (\textbf{D}), magnetic dipole (\textbf{m}), electric quadrupole $(\hat{\textbf{Q}})$, and magnetic quadrupole $(\hat{\textbf{M}})$ is given by,
\begin{equation}
 E_{sca} = \frac{k_0^2}{4\pi\epsilon_0}\bigg\{\Big[\textbf{n}\times\big[\textbf{D}\times \textbf{n}\big]\,\Big]\, + \frac{1}{v_d}[\textbf{m}\times \textbf{n}]\ + \frac{ik_d}{6}\Big[\textbf{n}\times \big[\textbf{n}\times \hat{\textbf{Q}}{n}\big]\,\Big]\, + \frac{ik_d}{2v_d}[\textbf{n}\times (\hat{\textbf{M}}{n})]\,\bigg\}
\end{equation}
where $\textbf{n}$ is the unit vector in the radial direction, $k_0$ is the wave number in vacuum, $k_d$ is the wave number and $v_d$ is the velocity of the light in the surrounding medium. The total electric dipole moment obtained from electric dipole moment $p$ and the toroidal dipole moment $\textbf{T}$ is defined by $\textbf{D} = \textbf{p} + i\frac{k_d}{v_d}\textbf{T}$. $\textbf{p}$ written in terms of displacement current density $\textbf{J}$ is $\textbf{p} = \frac{i}{\omega} \int \textbf{J}dr$, where $\textbf{J} = -i\omega\varepsilon_0[n^2 - 1]\textbf{E}$  and $\textbf{T} = \frac{1}{10c}\int [(\textbf{r}\cdot \textbf{J})r - 2r^2]dr$. We cite to the reference\cite{evlyukhin2016optical} for the definitions of higher order moments.

\subsection{Calculation of  $g^{(2)}(0)$}
The open-source Quantum Toolbox in Python (QuTiP)  \cite{johansson2012qutip} is utilized to analyze the quantum optical properties of the light–matter coupled \ch{WS2} metasurface system. A cavity quantum electrodynamics (cQED) framework was used to model the hybrid system by considering the exciton in the \ch{WS2} as a two-level system coupled to a quasi-BIC state which is considered as a quantized cavity mode. Lindblad-type collapse operators were used to account for dissipative phenomena including cavity photon loss and exciton decay. By solving the master equation, the steady-state density matrix ($\rho_{ss}$) has been computed. The second-order correlation function's quantum mechanical definition is

\begin{equation}
g^{(2)}(\tau) = \frac{\langle \hat{a}^\dagger(0)\hat{a}^\dagger(\tau)\hat{a}(\tau)\hat{a}(0) \rangle}{\langle \hat{a}^\dagger \hat{a} \rangle^2}
\end{equation}

with \( \hat{a} \) and \( \hat{a}^\dagger \) signifying the annihilation and creation operators, respectively. Normalized by the square of the average photon number, this formula measures the likelihood of finding a photon at time \( t + \tau \), conditional on a detection at time \( t \). This reduces at \( \tau = 0 \) to

\begin{equation}
g^{(2)}(0) = \frac{\langle \hat{a}^\dagger \hat{a}^\dagger \hat{a} \hat{a} \rangle}{\langle \hat{a}^\dagger \hat{a} \rangle^2}
\end{equation}

\( g^{(2)}(0) \), the scond order correlation function at zero delay, indicates the likelihood of finding two photons simultaneously or within a short time interval. In coherent light source, such as an ideal laser, emits photons separately, resulting in \( g^{(2)}(0) = 1 \). If the light source exhibits intensity fluctuations that result in photon bunching, as seen in thermal light from a blackbody, then \( g^{(2)}(0) > 1 \). If the light exhibits photon antibunching, then \( g^{(2)}(0) < 1 \), indicating that the source has quantum mechanical characteristics and is capable of emitting single photons \cite{paul1982photon, nogueira2001experimental}.

\textbf{Acknowledgements} \par 
We are thankful to Mahindra University, Hyderabad, for supporting and funding for our research.

\bibliographystyle{unsrtnat}
\bibliography{references}  






\end{document}